\RequirePackage{lineno}
\documentclass[aps,prd,twocolumn,superscriptaddress,showpacs,preprintnumbers,amsmath,amssymb,nofootinbib]{revtex4-2}

\newcommand{\RN}[1]{%
  \textup{\uppercase\expandafter{\romannumeral#1}}%
}

\usepackage{graphicx} 
\usepackage{dcolumn}  
\usepackage{epstopdf}
\usepackage{multirow}
\usepackage{color}
\usepackage[bookmarks]{hyperref}
\usepackage{amsmath}
\usepackage{nccmath}
\usepackage[symbol]{footmisc}

\graphicspath{{ps}}

\renewcommand{\arraystretch}{1.1}

\newcommand{\ra}{\rightarrow}

\newcommand{\GeV}{{\rm GeV}/c^{2}}
\newcommand{\MeV}{{\rm MeV}/c^{2}}

\newcommand{\K}{K}

\newcommand{\Kminus}{K^{-}}
\newcommand{\piminus}{\pi^{-}}
\newcommand{\piplus}{\pi^{+}}

\newcommand{\p}{p}

\newcommand{\lambdac}{\Lambda^{+}_{c}}

\newcommand{\elpi}{\lambdac \ra \eta \Lambda \piplus}
\newcommand{\pkpi}{\lambdac \ra \p \Kminus \piplus}
\newcommand{\sepi}{\lambdac \ra \eta \Sigma^{0} \piplus}
\newcommand{\el}{ \lambdac \ra \Lambda(1670) \piplus }
\newcommand{\lpi}{\lambdac \ra \eta \Sigma(1385)^{+} }

\newcommand{\belpi}{\mathcal{B}(\lambdac \ra \eta \Lambda \piplus)}
\newcommand{\bpkpi}{\mathcal{B}(\lambdac \ra \p \Kminus \piplus)}
\newcommand{\bsepi}{\mathcal{B}(\lambdac \ra \eta \Sigma^{0} \piplus)}
\newcommand{\bel}{\mathcal{B}(\lambdac \ra \Lambda(1670) \piplus) \times \mathcal{B}(\Lambda(1670) \ra \eta \Lambda) }
\newcommand{\blpi}{\mathcal{B}(\lambdac \ra \eta \Sigma(1385)^{+}) }

\newcommand{\resultelpiratio}{0.293 \pm 0.003 \pm 0.014}
\newcommand{\resultsepiratio}{0.120 \pm 0.006 \pm 0.006}
\newcommand{\resultelratio}{(5.54 \pm 0.29 \pm 0.73 ) \times 10^{-2}}
\newcommand{\resultlpiratio}{0.192 \pm 0.006 \pm 0.016}

\newcommand{\resultelpiabs}{(1.84 \pm 0.02 \pm 0.09 \pm 0.09)\%}
\newcommand{\resultsepiabs}{(7.56 \pm 0.39 \pm 0.37 \pm 0.39) \times 10^{-3}}
\newcommand{\resultelabs}{(3.48 \pm 0.19 \pm 0.46 \pm 0.18) \times 10^{-3}}
\newcommand{\resultlpiabs}{(1.21 \pm 0.04 \pm 0.10 \pm 0.06)\%}

\newcommand{\resultlm}{ 1674.3 \pm 0.8 \pm 4.9~\MeV}
\newcommand{\resultlw}{ 36.1 \pm 2.4 \pm 4.8~{\rm MeV} }
\newcommand{\resultsm}{ 1384.8 \pm 0.3 \pm 1.4~\MeV}
\newcommand{\resultsw}{ 38.1 \pm 1.5 \pm 2.1~{\rm MeV} }

\begin{document}



\title{ \quad\\[1.0cm] Measurement of Branching Fractions of $\elpi$, $\eta \Sigma^{0} \piplus$, $\Lambda(1670) \piplus$, and $\eta \Sigma(1385)^{+}$}

{\renewcommand{\thefootnote}{\fnsymbol{footnote}}}
\setcounter{footnote}{0}

\noaffiliation
\affiliation{University of the Basque Country UPV/EHU, 48080 Bilbao}
\affiliation{University of Bonn, 53115 Bonn}
\affiliation{Brookhaven National Laboratory, Upton, New York 11973}
\affiliation{Budker Institute of Nuclear Physics SB RAS, Novosibirsk 630090}
\affiliation{Faculty of Mathematics and Physics, Charles University, 121 16 Prague}
\affiliation{Chonnam National University, Gwangju 61186}
\affiliation{University of Cincinnati, Cincinnati, Ohio 45221}
\affiliation{Deutsches Elektronen--Synchrotron, 22607 Hamburg}
\affiliation{Duke University, Durham, North Carolina 27708}
\affiliation{University of Florida, Gainesville, Florida 32611}
\affiliation{Department of Physics, Fu Jen Catholic University, Taipei 24205}
\affiliation{Key Laboratory of Nuclear Physics and Ion-beam Application (MOE) and Institute of Modern Physics, Fudan University, Shanghai 200443}
\affiliation{Gifu University, Gifu 501-1193}
\affiliation{II. Physikalisches Institut, Georg-August-Universit\"at G\"ottingen, 37073 G\"ottingen}
\affiliation{SOKENDAI (The Graduate University for Advanced Studies), Hayama 240-0193}
\affiliation{Gyeongsang National University, Jinju 52828}
\affiliation{Department of Physics and Institute of Natural Sciences, Hanyang University, Seoul 04763}
\affiliation{University of Hawaii, Honolulu, Hawaii 96822}
\affiliation{High Energy Accelerator Research Organization (KEK), Tsukuba 305-0801}
\affiliation{J-PARC Branch, KEK Theory Center, High Energy Accelerator Research Organization (KEK), Tsukuba 305-0801}
\affiliation{Higher School of Economics (HSE), Moscow 101000}
\affiliation{Forschungszentrum J\"{u}lich, 52425 J\"{u}lich}
\affiliation{IKERBASQUE, Basque Foundation for Science, 48013 Bilbao}
\affiliation{Indian Institute of Science Education and Research Mohali, SAS Nagar, 140306}
\affiliation{Indian Institute of Technology Bhubaneswar, Satya Nagar 751007}
\affiliation{Indian Institute of Technology Guwahati, Assam 781039}
\affiliation{Indian Institute of Technology Hyderabad, Telangana 502285}
\affiliation{Indian Institute of Technology Madras, Chennai 600036}
\affiliation{Indiana University, Bloomington, Indiana 47408}
\affiliation{Institute of High Energy Physics, Chinese Academy of Sciences, Beijing 100049}
\affiliation{Institute of High Energy Physics, Vienna 1050}
\affiliation{Institute for High Energy Physics, Protvino 142281}
\affiliation{INFN - Sezione di Napoli, 80126 Napoli}
\affiliation{INFN - Sezione di Torino, 10125 Torino}
\affiliation{Advanced Science Research Center, Japan Atomic Energy Agency, Naka 319-1195}
\affiliation{J. Stefan Institute, 1000 Ljubljana}
\affiliation{Institut f\"ur Experimentelle Teilchenphysik, Karlsruher Institut f\"ur Technologie, 76131 Karlsruhe}
\affiliation{Kavli Institute for the Physics and Mathematics of the Universe (WPI), University of Tokyo, Kashiwa 277-8583}
\affiliation{Kennesaw State University, Kennesaw, Georgia 30144}
\affiliation{Department of Physics, Faculty of Science, King Abdulaziz University, Jeddah 21589}
\affiliation{Korea Institute of Science and Technology Information, Daejeon 34141}
\affiliation{Korea University, Seoul 02841}
\affiliation{Kyoto Sangyo University, Kyoto 603-8555}
\affiliation{Kyungpook National University, Daegu 41566}
\affiliation{Universit\'{e} Paris-Saclay, CNRS/IN2P3, IJCLab, 91405 Orsay}
\affiliation{P.N. Lebedev Physical Institute of the Russian Academy of Sciences, Moscow 119991}
\affiliation{Liaoning Normal University, Dalian 116029}
\affiliation{Faculty of Mathematics and Physics, University of Ljubljana, 1000 Ljubljana}
\affiliation{Ludwig Maximilians University, 80539 Munich}
\affiliation{Luther College, Decorah, Iowa 52101}
\affiliation{University of Maribor, 2000 Maribor}
\affiliation{Max-Planck-Institut f\"ur Physik, 80805 M\"unchen}
\affiliation{School of Physics, University of Melbourne, Victoria 3010}
\affiliation{University of Mississippi, University, Mississippi 38677}
\affiliation{University of Miyazaki, Miyazaki 889-2192}
\affiliation{Moscow Physical Engineering Institute, Moscow 115409}
\affiliation{Graduate School of Science, Nagoya University, Nagoya 464-8602}
\affiliation{Universit\`{a} di Napoli Federico II, 80126 Napoli}
\affiliation{Nara Women's University, Nara 630-8506}
\affiliation{National Central University, Chung-li 32054}
\affiliation{National United University, Miao Li 36003}
\affiliation{Department of Physics, National Taiwan University, Taipei 10617}
\affiliation{H. Niewodniczanski Institute of Nuclear Physics, Krakow 31-342}
\affiliation{Nippon Dental University, Niigata 951-8580}
\affiliation{Niigata University, Niigata 950-2181}
\affiliation{University of Nova Gorica, 5000 Nova Gorica}
\affiliation{Novosibirsk State University, Novosibirsk 630090}
\affiliation{Osaka City University, Osaka 558-8585}
\affiliation{Pacific Northwest National Laboratory, Richland, Washington 99352}
\affiliation{Panjab University, Chandigarh 160014}
\affiliation{Peking University, Beijing 100871}
\affiliation{University of Pittsburgh, Pittsburgh, Pennsylvania 15260}
\affiliation{Research Center for Nuclear Physics, Osaka University, Osaka 567-0047}
\affiliation{Meson Science Laboratory, Cluster for Pioneering Research, RIKEN, Saitama 351-0198}
\affiliation{RIKEN BNL Research Center, Upton, New York 11973}
\affiliation{Department of Modern Physics and State Key Laboratory of Particle Detection and Electronics, University of Science and Technology of China, Hefei 230026}
\affiliation{Seoul National University, Seoul 08826}
\affiliation{Showa Pharmaceutical University, Tokyo 194-8543}
\affiliation{Soochow University, Suzhou 215006}
\affiliation{Soongsil University, Seoul 06978}
\affiliation{Sungkyunkwan University, Suwon 16419}
\affiliation{School of Physics, University of Sydney, New South Wales 2006}
\affiliation{Department of Physics, Faculty of Science, University of Tabuk, Tabuk 71451}
\affiliation{Tata Institute of Fundamental Research, Mumbai 400005}
\affiliation{Department of Physics, Technische Universit\"at M\"unchen, 85748 Garching}
\affiliation{School of Physics and Astronomy, Tel Aviv University, Tel Aviv 69978}
\affiliation{Toho University, Funabashi 274-8510}
\affiliation{Earthquake Research Institute, University of Tokyo, Tokyo 113-0032}
\affiliation{Department of Physics, University of Tokyo, Tokyo 113-0033}
\affiliation{Tokyo Institute of Technology, Tokyo 152-8550}
\affiliation{Tokyo Metropolitan University, Tokyo 192-0397}
\affiliation{Virginia Polytechnic Institute and State University, Blacksburg, Virginia 24061}
\affiliation{Wayne State University, Detroit, Michigan 48202}
\affiliation{Yamagata University, Yamagata 990-8560}
\affiliation{Yonsei University, Seoul 03722}

  \author{J.~Y.~Lee}\affiliation{Seoul National University, Seoul 08826} 
  \author{K.~Tanida}\affiliation{Advanced Science Research Center, Japan Atomic Energy Agency, Naka 319-1195} 
  \author{Y.~Kato}\affiliation{Graduate School of Science, Nagoya University, Nagoya 464-8602} 
  \author{S.~K.~Kim}\affiliation{Seoul National University, Seoul 08826} 
  \author{S.~B.~Yang}\thanks{Corresponding author. sbyang@korea.ac.kr}\affiliation{Korea University, Seoul 02841} 
   
  \author{I.~Adachi}\affiliation{High Energy Accelerator Research Organization (KEK), Tsukuba 305-0801}\affiliation{SOKENDAI (The Graduate University for Advanced Studies), Hayama 240-0193} 
  \author{J.~K.~Ahn}\affiliation{Korea University, Seoul 02841} 
  \author{H.~Aihara}\affiliation{Department of Physics, University of Tokyo, Tokyo 113-0033} 
  \author{S.~Al~Said}\affiliation{Department of Physics, Faculty of Science, University of Tabuk, Tabuk 71451}\affiliation{Department of Physics, Faculty of Science, King Abdulaziz University, Jeddah 21589} 
  \author{D.~M.~Asner}\affiliation{Brookhaven National Laboratory, Upton, New York 11973} 
% ...
\collaboration{The Belle Collaboration}

  
\begin{abstract}
  We report branching fraction measurements of four decay modes of the $\lambdac$ baryon, each of which includes an $\eta$ meson and a $\Lambda$ baryon in the final state, and all of which are measured relative to the $\pkpi$ decay mode. 
  The results are based on a $980~\mathrm{fb^{-1}}$ data sample collected by the Belle detector at the KEKB asymmetric-energy $e^{+}e^{-}$ collider.
  Two decays, $\sepi$ and $\Lambda(1670)\piplus$, are observed for the first time, while the measurements of the other decay modes, $\elpi$ and $\eta\Sigma(1385)^{+}$, are more precise than those made previously.
 We obtain $\belpi/\bpkpi$ = $\resultelpiratio$, $\bsepi/\bpkpi$ = $\resultsepiratio$, $\bel/\bpkpi$ = $\resultelratio$, and $\blpi/\bpkpi$ = $\resultlpiratio$. 
 The mass and width of the $\Lambda(1670)$ are also precisely determined to be $\resultlm$ and $\resultlw$, respectively, where the uncertainties are statistical and systematic, respectively.
\end{abstract}


\maketitle

\tighten


\section{Introduction}

The branching fractions of weakly decaying charmed baryons provide a way to study both strong and weak interactions.
Although there are theoretical models that estimate the branching fractions, for example constituent quark models and Heavy Quark Effective Theories (HQET)~\cite{br_model1,br_model2}, the lack of experimental measurements of branching fractions of charmed baryons makes it difficult to test the models.
Therefore, branching fraction measurements of new decay modes of the $\lambdac$ or known decay modes with higher statistics are crucial.
Model-independent measurements of the branching fraction of $\pkpi$ by Belle~\cite{pkpi_belle} and BES\RN{3}~\cite{pkpi_bes} now enable branching ratios measured relative to the $\pkpi$ mode to be converted to absolute branching fraction measurements with high precision~\cite{charge-conjugate}.
The $\elpi$ decay mode is especially interesting since it has been suggested~\cite{l1670_a0} that it is an ideal decay mode to study the $\Lambda(1670)$ and $a_{0}(980)$ because, for any combination of two particles in the final state, the isospin is fixed.

Two different models have been proposed to explain the structure of the $\Lambda(1670)$.
One is based on a quark model and assigns it to be the SU(3) octet partner of the $N(1535)$~\cite{l1670_quarkmodel}.
The other describes the $\Lambda(1670)$ as a $K\Xi$ bound state using a meson-baryon model that has also been used to describe the $\Lambda(1405)$ as a $\bar{K}N$ bound state \cite{l1670_mesonbaryon}.
There have been few experimental efforts to confirm the structure of the $\Lambda(1670)$; and the interpretation of partial-wave analyses of $\bar{K}N$ scattering data depends on theoretical models~\cite{l1670_zhang, l1670_kamano}.
Here we investigate the production and decays of the $\Lambda(1670)$ in the resonant substructure of the $\elpi$ decay, in order to elucidate the nature of this particle.

We present measurements of branching fractions for the four decay modes, $\elpi$, $\sepi$, $\el$, and $\lpi$, all measured relative to the $\pkpi$ decay mode.
The branching fraction of the $\el$ decay mode is given as the product $\mathcal{B}(\el) \times \mathcal{B}(\Lambda(1670) \ra \eta\Lambda)$, because $\mathcal{B}(\Lambda(1670) \ra \eta\Lambda)$ is not well-determined.
The $\el$ and $\lpi$ decay modes are studied as resonant structures in the $\elpi$ decay, while the $\sepi$ decay is observed indirectly as a feed-down to the $M(\eta\Lambda\piplus)$ spectrum.
While $\mathcal{B}(\elpi)$ and $\mathcal{B}(\lpi)$ have previously been measured by CLEO~\cite{CLEO_br} and by BES\RN{3}~\cite{BES_br}, we report the first observation of the $\sepi$ and $\el$ decay modes and their branching fractions.
We also make precise measurements of the masses and widths of the $\Lambda(1670)$ and $\Sigma(1385)^{+}$.

\section{Data Sample and Monte Carlo Simulation}

This measurement is based on data recorded at or near the $\Upsilon(1S)$, $\Upsilon(2S)$, $\Upsilon(3S)$, $\Upsilon(4S)$, and $\Upsilon(5S)$ resonances by the Belle detector at the KEKB asymmetric-energy $e^{+}e^{-}$ collider~\cite{KEKB_1}.
The total data sample corresponds to an integrated luminosity of $980~\mathrm{fb^{-1}}$.
The Belle detector is a large-solid-angle magnetic spectrometer that consists of a silicon vertex detector (SVD), a central drift chamber (CDC), an array of aerogel threshold Cherenkov counters (ACC), a barrel-like arrangement of time-of-flight scintillation counters (TOF), and an electromagnetic calorimeter comprising CsI(Tl) crystals (ECL) located inside a superconducting solenoid coil that provides a 1.5~T magnetic field.
An iron flux-return located outside of the coil is instrumented to detect $K^{0}_{L}$ mesons and to identify muons.
The detector is described in detail elsewhere~\cite{Belle_1}.
Two inner detector configurations were used.
A 2.0-cm radius beampipe and a three-layer silicon vertex detector were used for the first sample of 156 $\mathrm{fb^{-1}}$, while a 1.5-cm radius beampipe, a four-layer silicon detector and a small-cell inner drift chamber were used to record the remaining 824 $\mathrm{fb^{-1}}$.

Monte Carlo (MC) simulation events are generated with PYTHIA~\cite{pythia} and EvtGen~\cite{evtgen_1} and propagated by GEANT3~\cite{geant3}.
The effect of final-state radiation is taken into account in the simulation using the PHOTOS~\cite{photos} package.
A generic MC simulation sample, having the same integrated luminosity as real data, is used to optimize selection criteria for $\elpi$ signal events.
We also generate several signal MC simulation samples of specific $\Lambda_{c}^{+}$ decays in order to study particle reconstruction efficiencies and the detector performance; the signal MC events follow a uniform distribution in phase space.

\section{Event Selection}

 We reconstruct $\lambdac$ candidates via $\elpi$ decays with the $\eta$ and $\Lambda$ in $\eta \ra \gamma \gamma$ and $\Lambda \ra \p \piminus$ decays.
Starting from selection criteria typically used in other charmed-hadron analyses at Belle \cite{sb_dcs,ew_cp}, our final criteria are determined through a figure-of-merit (FoM) study based on the generic MC sample.
We optimize the FoM, defined as $n_{\rm sig} /\sqrt{n_{\rm sig}+n_{\rm bkg}}$, where $n_{\rm sig}$ is the number of reconstructed $\Lambda_{c}^{+}$ signal events while $n_{\rm bkg}$ is the number of background events.
The yields $n_{\rm sig}$ and $n_{\rm bkg}$ are counted in the $M(\eta\Lambda\piplus)$ range from $2.2755~\GeV$ to $2.2959~\GeV$.

 The $\eta$ meson candidates are reconstructed from photon pairs in which $M(\gamma\gamma)$ is in the range $0.50$-$0.58~\GeV$ corresponding to an efficiency of about $79\%$.
 A mass-constrained fit is performed to improve the momentum resolution of $\eta$ candidates, and the fitted momentum and energy are used for the subsequent steps of analysis.
In addition, we require $\eta$ candidates to have momenta greater than 0.4 ${\rm GeV}/c$ and an energy asymmetry, defined as $|(E(\gamma_{1})-E(\gamma_{2}))/(E(\gamma_{1})+E(\gamma_{2}))|$, less than 0.8.
For the selection of photons, the energy deposited in the ECL is required to be greater than 50 $\rm MeV$ for the barrel region and greater than 100 $\rm MeV$ for the endcap region~\cite{Belle_1}.
In order to reject neutral hadrons, the ratio between energy deposited in the $3 \times 3$ array of crystals centered on the crystal with the highest energy, to that deposited in the corresponding $5 \times 5$ array of crystals, is required to be greater than 0.85.
To reduce the background in the $\eta$ signal region due to photons from $\pi^{0}$ decays, the photons used to reconstruct the $\eta$ candidates are not allowed to be a part of a reconstructed $\pi^{0}$ with mass between 0.12 $\GeV$ and 0.15 $\GeV$.

 Charged $\pi^{+}$ candidates are selected using requirements on a distance-of-closest-approach (DOCA) to the interaction point (IP) of less than 2.0 cm in the beam direction $(z)$ and less than 0.2 cm in the transverse $(r)$ direction.
Measurements from CDC, TOF, and ACC are combined to form particle identification (PID) likelihoods $\mathcal{L}(h)$ ($h =$ $p^{\pm}$, $K^{\pm}$, or $\pi^{\pm}$), and the $\mathcal{L}(h:h^{'})$, defined as $\mathcal{L}(h)/[\mathcal{L}(h) + \mathcal{L}(h^{'})]$, is the ratio of likelihoods for $h$ and $h^{'}$.
For the selection of $\pi^{+}$, $\mathcal{L}(\pi:\K) > 0.2$ and $\mathcal{L}(\pi:p) > 0.4$ are required.
Furthermore, the electron likelihood ratio $\mathcal{R}(e)$, derived from ACC, CDC, and ECL measurements~\cite{pid_e}, is required to be less than 0.7.

 We reconstruct $\Lambda$ candidates via $\Lambda \ra \p \piminus$ decays in the mass range, $1.108~\GeV < M(p\piminus) < 1.124~\GeV$, and selected using $\Lambda$-momentum-dependent criteria based on four parameters: the distance between two daughter tracks along the $z$ direction at their closest approach; the minimum distance between daughter tracks and the IP in the transverse plane; the angular difference between the $\Lambda$ flight direction and the direction pointing from the IP to the $\Lambda$ decay vertex in the transverse plane; and the flight length of $\Lambda$ in the transverse plane.
We require $\mathcal{L}(\p:\pi) > 0.6$ for the proton from the $\Lambda$ decay.

Finally, $\eta$, $\Lambda$, and $\piplus$ candidates are combined to form a $\lambdac$ with its daughter tracks fitted to a common vertex.
The $\chi^{2}$ value from the vertex fit is required to be less than 40, with an efficiency of about 87$\%$.
To reduce combinatorial background, especially from $B$ meson decays, the scaled momentum $x_{p} = p^{*}/p_{\text{max}}$ is required to be greater than 0.51; here, $p^{*}$ is the momentum of $\lambdac$ in the center-of-mass frame and $p_{\text{max}}$ is the maximum possible momentum. 

Since the branching fractions are determined relative to $\mathcal{B}(\pkpi)$, $\lambdac$ candidates from $\pkpi$ decays are also reconstructed using the same selection criteria in Ref.~\cite{sb_dcs} except for the scaled momentum requirement of the $\lambdac$, which is chosen to be the same as that used for the $\elpi$ channel. 
All charged tracks in the $\pkpi$ decay are required to have their DOCA less than 2.0 $\rm cm$ and 0.1 $\rm cm$ in the $z$ and $r$ directions, respectively, and at least one SVD hit in both the $z$ and $r$ directions.
The PID requirements are $\mathcal{L}(p:K) > 0.9$ and $\mathcal{L}(p:\pi) > 0.9$ for $p$, $\mathcal{L}(K:p) > 0.4$ and $\mathcal{L}(K:\pi) > 0.9$ for $K$, and $\mathcal{L}(\pi:p) > 0.4$ and $\mathcal{L}(\pi:K) > 0.4$ for $\pi$. 
In addition, $\mathcal{R}(e) < 0.9$ is required for all tracks.
The charged tracks from the $\Lambda_{c}^{+}$ decay are fitted to a common vertex and the $\chi^{2}$ value from the vertex fit must be less than 40.

\section{ Branching Fractions of $\elpi$ and $\eta\Sigma^{0}\piplus$ Modes }
\label{sec:branching}
The branching fractions of the $\elpi$ and $\eta\Sigma^{0}\piplus$ decays are calculated relative to that of the $\pkpi$ decay using the efficiency-corrected event yields via the following equation,
\begin{equation}  \label{eq:branching_fraction}
  \begin{aligned}
    \frac{\mathcal{B}({\rm Decay~}{\rm Mode})}{\mathcal{B}(\pkpi)} = \frac{y({\rm Decay~Mode})}{\mathcal{B}_{\rm PDG}\times y(\pkpi)},
    \end{aligned}
\end{equation}
where Decay Mode is either $\elpi$ or $\sepi$, and $y({\rm Decay~Mode})$ refers to the efficiency-corrected yield of the corresponding decay mode. 
Here $\mathcal{B}_{\rm PDG}$ denotes subdecay branching fractions of the $\eta$, $\Lambda$, and $\Sigma^{0}$; we use $\mathcal{B}(\eta \rightarrow \gamma \gamma)=(39.41 \pm 0.20)\%$, $\mathcal{B}(\Lambda \rightarrow p \pi^{-})=(63.9 \pm 0.5)\%$, and $\mathcal{B}(\Sigma^{0} \rightarrow \Lambda \gamma)=100\%$ from Ref.~\cite{pdg}.

\begin{figure}[t]
  \includegraphics[width=1.0\linewidth]{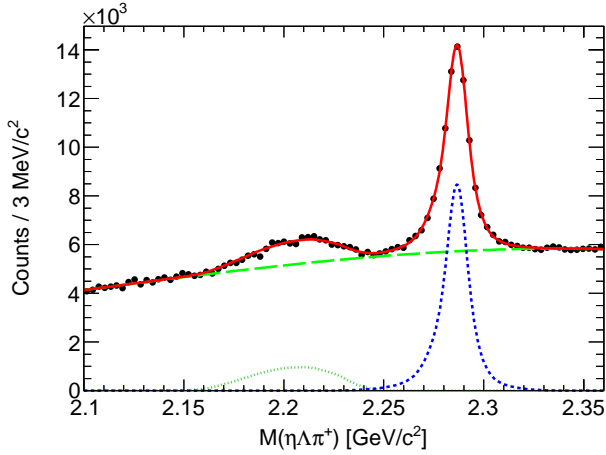}
  \caption
      {
        Fit to the $M(\eta\Lambda\piplus)$ distribution. The curves indicate the fit result: the total PDF (solid red), signal from $\sepi$ channel with a missing photon from the $\Sigma^{0}$ decay (dotted dark green), signal from $\elpi$ decay (dashed blue) and combinatorial backgrounds (long-dashed green).
      }
      \label{fig:data_elpi_all}
\end{figure}

\begin{figure}[t]
  \begin{minipage}[!htb]{1.0\linewidth}
    \includegraphics[height=0.75\textwidth,width=1.0\textwidth]{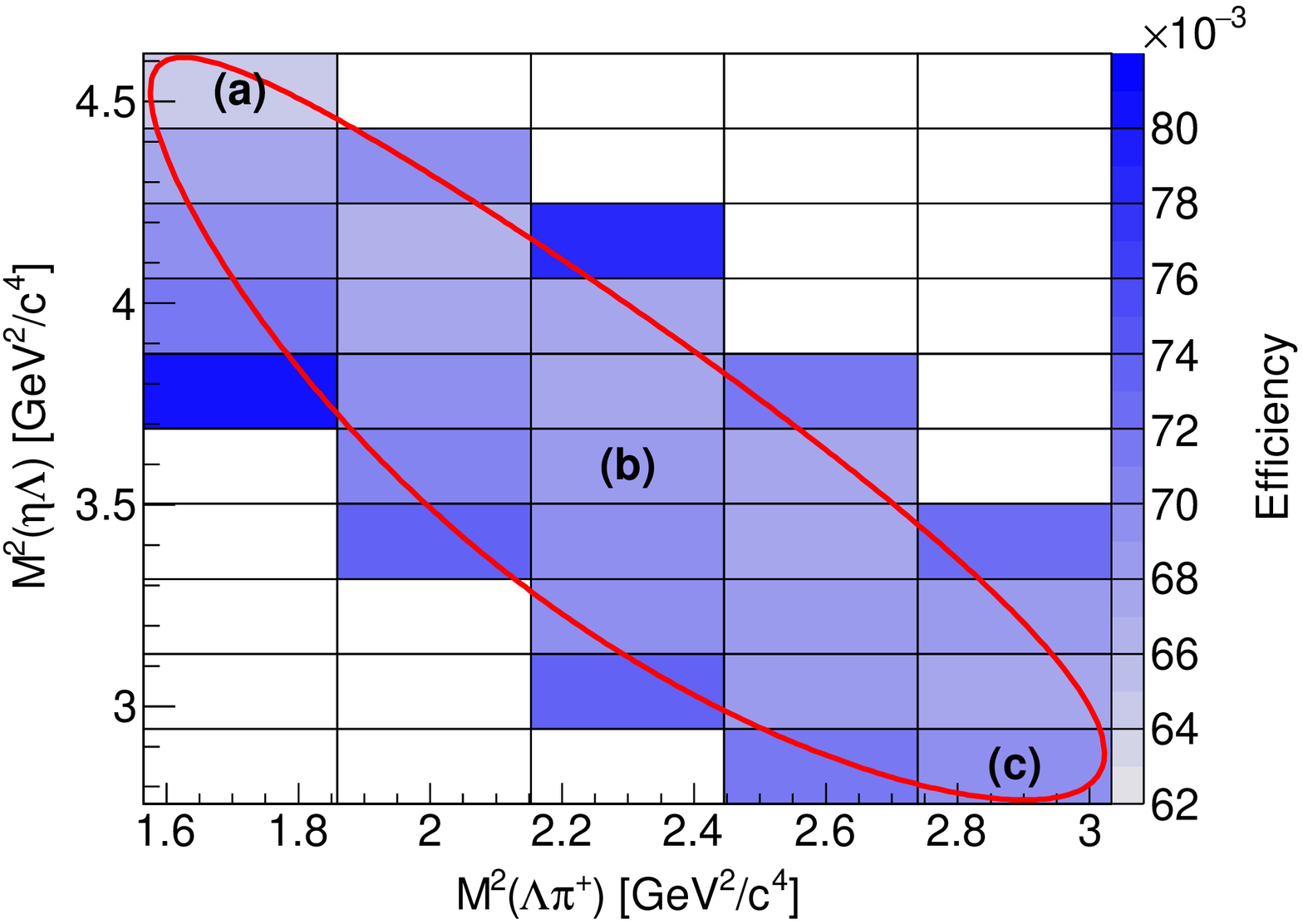}
  \end{minipage}
  \begin{minipage}[!htb]{1.0\linewidth}
    \includegraphics[height=0.75\textwidth,width=1.0\textwidth]{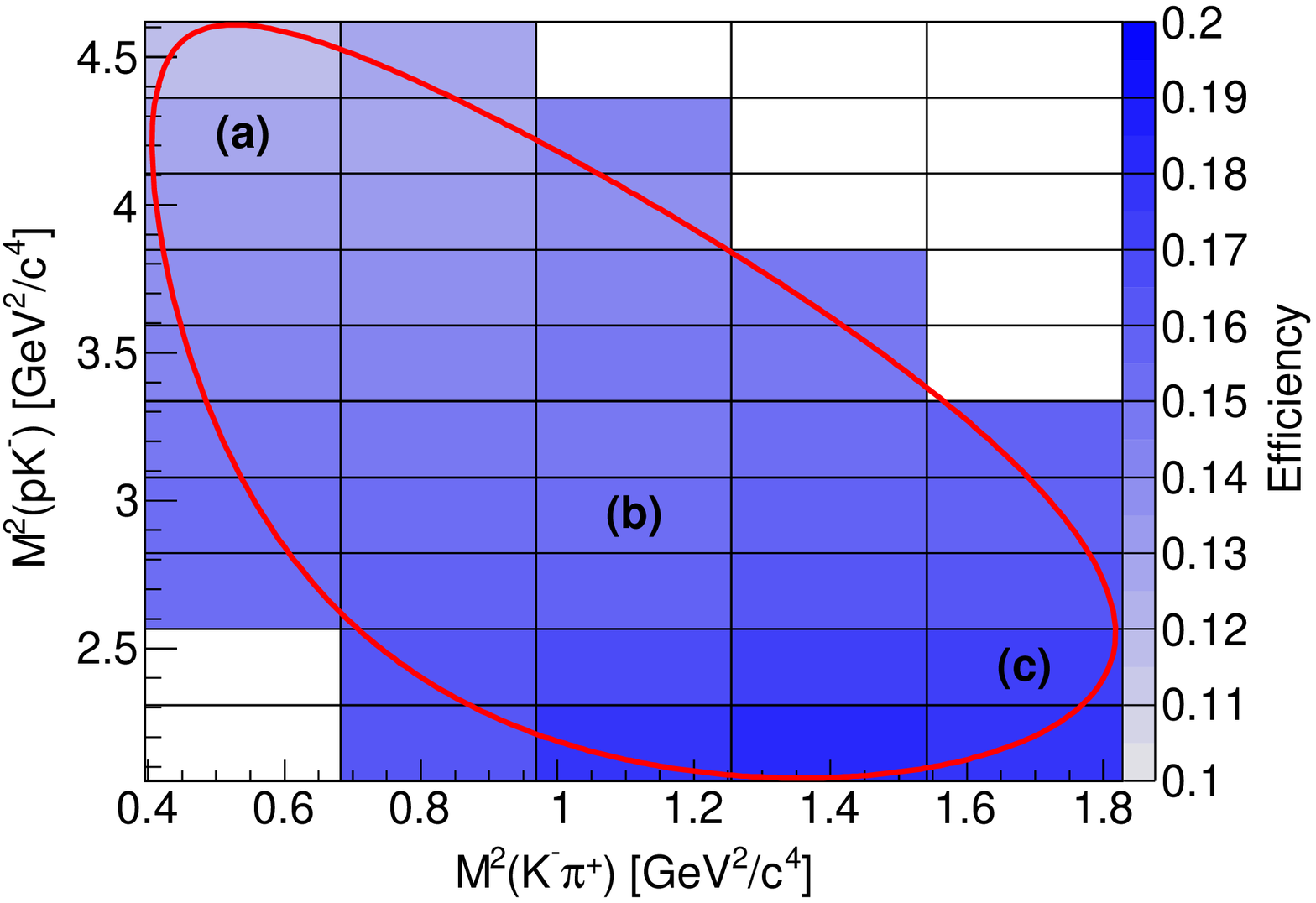}
  \end{minipage}\hspace{\fill}
  \caption
      {
        Distribution of the reconstruction efficiencies over the Dalitz plots divided into the $10 \times 5$ bins of $M^{2}(\Lambda\piplus)$ vs $M^2(\eta\Lambda)$ for the $\elpi$ channel (top) and of $M^{2}(\Kminus\piplus)$ vs $M^{2}(p\Kminus)$ for the $\pkpi$ channel (bottom).
        The red lines indicate the Dalitz plot boundaries.
        The fits in the three sample bins of (a), (b) and (c) are shown in Fig.~\ref{fig:example_elpi}. for the $\elpi$ channel and in Fig.~\ref{fig:example_pkpi} for the $\pkpi$ channel.
      }
      \label{fig:eff_dalitz}
\end{figure}

Figure~\ref{fig:data_elpi_all} shows the $M(\eta\Lambda\piplus)$ spectrum after the event selection described in the previous section.
In the spectrum, we find a peaking structure from the $\lambdac \ra \eta \Lambda \piplus$ channel at $2.286~{\rm GeV}/c^{2}$. 
The enhancement to the left of the peak corresponds to the $\sepi$ channel with a missing photon from the $\Sigma^{0} \rightarrow \Lambda\gamma$ decay.   
First, we perform a binned-$\chi^{2}$ fit to the $M(\eta\Lambda\piplus)$ distribution to extract the $\sepi$ signal yield.
The probability density functions (PDFs) of the signals are modeled empirically based on MC samples as the sum of a Gaussian and two bifurcated Gaussian functions with a common mean for $\elpi$, and a histogram PDF for the feed-down of the $\sepi$ decay.
The latter PDF is derived from $\lambdac \ra \eta \Sigma^{0} \piplus$; $\Sigma^{0} \ra \Lambda\gamma$ decays where the photon decaying from the $\Sigma^{0}$ is not reconstructed.
The PDF of the combinatorial backgrounds used for the fit is a third-order polynomial function.
The signal yield for the feed-down from the $\sepi$ channel shown in Fig.~\ref{fig:data_elpi_all} is $17058 \pm 871$.
This yield is then corrected for the reconstruction efficiency obtained from MC to give an efficiency-corrected yield of $(3.05 \pm 0.16)\times10^{5}$, where the uncertainty is statistical only.
\par
\begin{figure}[t]
  \includegraphics[width=0.9\linewidth]{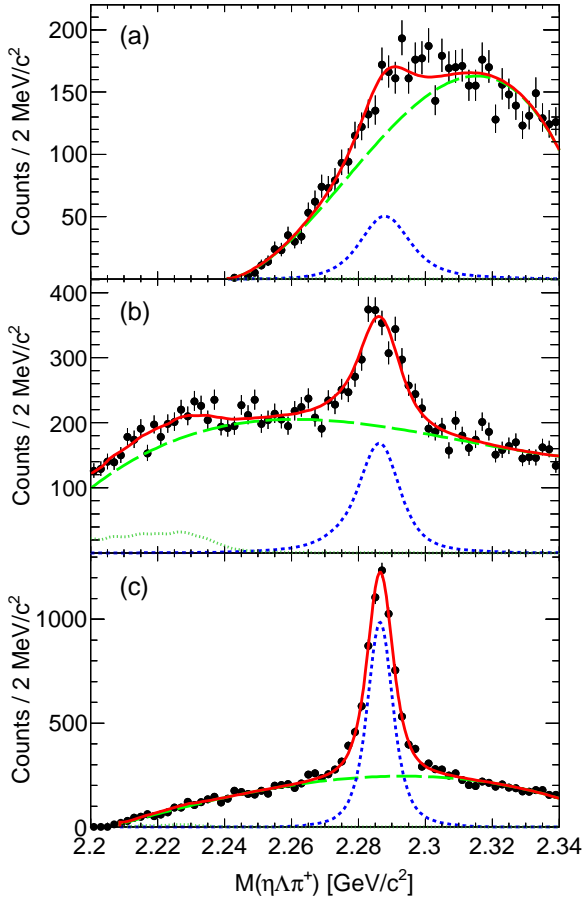}
  \caption
      {
        Fits in three sample Dalitz plot bins (see Fig.~\ref{fig:eff_dalitz}) of the $\elpi$ channel.
        The curves indicate the fit results: the total PDF (solid red), signal from the $\sepi$ channel with a missing photon from the $\Sigma^{0}$ decay (dotted dark green), signal from the $\elpi$ decay (dashed blue) and combinatorial backgrounds (long-dashed green).
      }
      \label{fig:example_elpi}
\end{figure}
\begin{figure}[t]
  \includegraphics[width=0.9\linewidth]{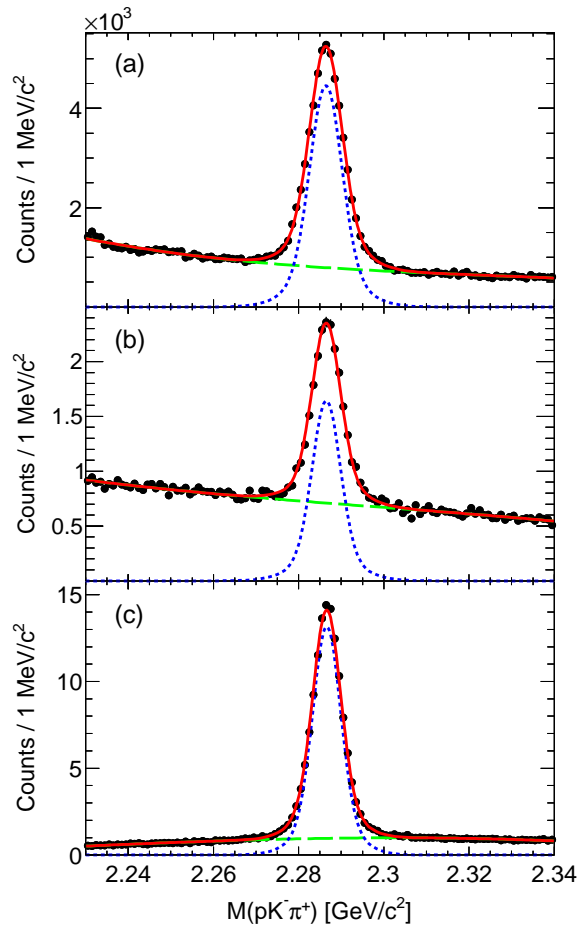}
  \caption
      {
        Fits in three sample Dalitz plot bins (see Fig.~\ref{fig:eff_dalitz}) of the $\pkpi$ channel.
        The curves indicate the fit results: the total PDF (solid red), signal from the $\pkpi$ decays (dashed blue) and combinatorial backgrounds (long-dashed green).
      }
      \label{fig:example_pkpi}
\end{figure}

On the other hand, the $\elpi$ and $p\Kminus\piplus$ channels have sufficiently large statistics to perform the yield extractions in individual bins of the Dalitz plot, in order to take into account the bin-to-bin variations of the efficiencies.
Figure~\ref{fig:eff_dalitz} shows the binning and the efficiencies over the Dalitz plots for $\elpi$ and $p\Kminus\piplus$, respectively.
For the fit to each bin of the $\elpi$ Dalitz plot, we use PDFs of the same form described above.
In the $p\Kminus\piplus$ channel, two Gaussian functions sharing a common mean value and a third-order polynomial function are used to represent the $p\Kminus\piplus$ signals and combinatorial backgrounds, respectively.
For the signal PDFs in both $\elpi$ and $p\Kminus\piplus$ fits, all parameters except for normalizations are fixed for each bin.
The fixed parameters are first obtained for each bin according to an MC simulation and later corrected by taking into account the difference of the fit results between data and MC samples over the entire region of the Dalitz plot.
For the fit to $\elpi$, all the parameters for the PDF attributed to the feed-down from the $\sepi$ decay with one photon missing are fixed, including the normalization based on the measured yield in this analysis.
The polynomial functions for the combinatorial backgrounds are floated for both $\elpi$ and $p\Kminus\piplus$ decays.
Figures~\ref{fig:example_elpi} and \ref{fig:example_pkpi} show examples of fits for three Dalitz plot bins.
For the $\elpi$ and $p\Kminus\piplus$ channels, the extracted yields are efficiency-corrected in each bin and summed up over the Dalitz plots. The results for the total efficiency-corrected signal yields are summarized in Table~\ref{tbl:yields}. 


\begin{table}[b]
  \caption
      {\label{tbl:yields}
        Summary of the efficiency-corrected signal yields for the various $\lambdac$ decay modes.
        The uncertainties are statistical.
        Note that for the $\elpi$ and $\pkpi$ decays, the signal yields are corrected in each Dalitz plot bin and summed, unlike the other decays. 
      }
  \begin{ruledtabular}
    \begin{tabular}{lcc}
      \multirow{2}{*}{\textrm{Decay modes}}&
      \multirow{2}{*}{\textrm{Extracted yields}}&
      \textrm{Efficiency-corrected}\\
       & & \textrm{yields} $[\times 10^{3}]$\\
      
      \colrule
      $\elpi$ & 51276 $\pm$ 454 & 741 $\pm$ 7\\
      $\pkpi$ & 1544580 $\pm$ 1552 & 10047 $\pm$ 10\\
      $\sepi$ & 17058 $\pm$ 871 & 305 $\pm$ 16 \\
      $\el$ & 9760 $\pm$ 519 & 140 $\pm$ 7 \\
      $\lpi$ & 29372 $\pm$ 875 & 423 $\pm$ 13\\
    \end{tabular}
  \end{ruledtabular}
\end{table}

Finally, we calculate the branching fractions using the efficiency-corrected signal yields and Eq.~\eqref{eq:branching_fraction}. The branching fractions are summarized in Table~\ref{tbl:branching_fraction}.

\begin{table}[!htb]
  \caption
      {\label{tbl:branching_fraction}
        Summary of the branching fractions for the various $\lambdac$ decay modes relative to the $\pkpi$ mode.
        The quoted uncertainties are statistical and systematic, respectively.
      }
  \begin{ruledtabular}
    \begin{tabular}{lc}
      \textrm{Decay modes}&
      $B({\rm Decay~Mode})/\bpkpi$\\
      \colrule
      $\elpi$ & $\resultelpiratio$ \\
      $\sepi$ & $\resultsepiratio$ \\
      $\el;$ &  \multirow{2}{*}{$\resultelratio$} \\
      $\Lambda(1670)\rightarrow\eta\Lambda$ & \\
      $\lpi$ &  $\resultlpiratio$ \\

    \end{tabular}
  \end{ruledtabular}
\end{table}

\section{ Analysis for Intermediate $\lambdac \ra \Lambda(1670)\piplus$ and $\eta\Sigma(1385)^{+}$ Modes }

\begin{figure}[t]
  \includegraphics[width=1.0\linewidth]{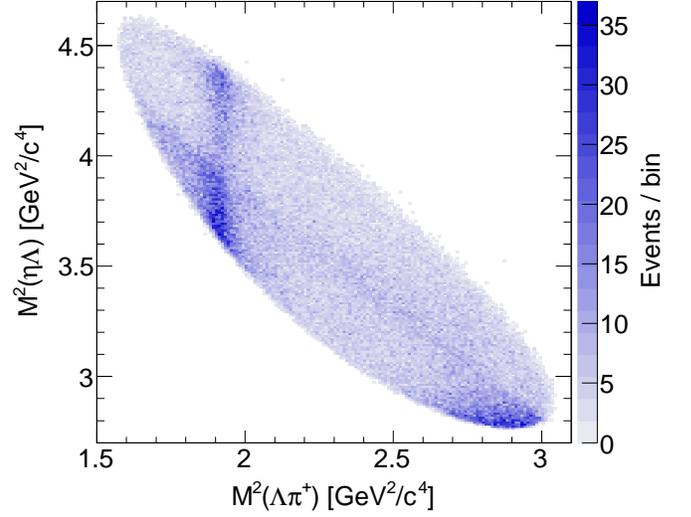}
  \caption
      { 
        Dalitz plot, invariant mass squared of $\Lambda\pi^{+}$ versus $\eta\Lambda$, for the $\elpi$ channel within $2.278~\GeV < M(\eta\Lambda\pi^{+}) < 2.294~\GeV$ in data sample. 
        Both bin widths of x and y axes are $0.01~{\rm GeV^{2}}/c^{4}$.
        Over the Dalitz plot, 48$\%$ of events are non-$\Lambda_{c}^{+}$ events.
        Horizontal and vertical bands at $M^{2}(\eta\Lambda)=2.79~{\rm GeV^{2}}/c^{4}$ and $M^{2}(\Lambda\pi^{+})=1.92~{\rm GeV^{2}}/c^{4}$ correspond to $\Lambda(1670)\pi^{+}$ and $\eta\Sigma(1385)^{+}$ subchannels, respectively.
      }
      \label{fig:data_dalitz}
\end{figure}
 Bands corresponding to $\Lambda_{c}^{+} \rightarrow \Lambda(1670)\pi^{+}$ and $\eta\Sigma(1385)^{+}$ resonant subchannels are visible on the Dalitz plot of $M^{2}(\Lambda\pi^{+})$ versus $M^{2}(\eta\Lambda$), shown in Fig.~\ref{fig:data_dalitz}. 
We also calculate the branching fractions of $\el$ and $\lpi$ decays using Eq.~\eqref{eq:branching_fraction}. 
In this case, ``Decay Mode" refers to $\el\ra\eta\Lambda\piplus$ or $\lpi$. 
For the $\lpi$ decay, $\mathcal{B}_{\rm PDG}$ includes the subdecay branching fraction of $\Sigma(1385)^{+} \ra \Lambda\pi^{+}$, $\mathcal{B}(\Sigma(1385)^{+} \ra \Lambda\pi^{+})=87.0 \pm 1.5\%$~\cite{pdg}. 
However, in the case of the $\el$, the subdecay branching fraction of $\Lambda(1670) \ra \eta\Lambda$ is not included because of its large uncertainty~\cite{pdg}. \par

\begin{figure}[!htb]
  \vspace{-0.0\linewidth}
  \begin{minipage}[!htb]{1.0\linewidth}
      \includegraphics[height=0.75\textwidth,width=1.0\textwidth]{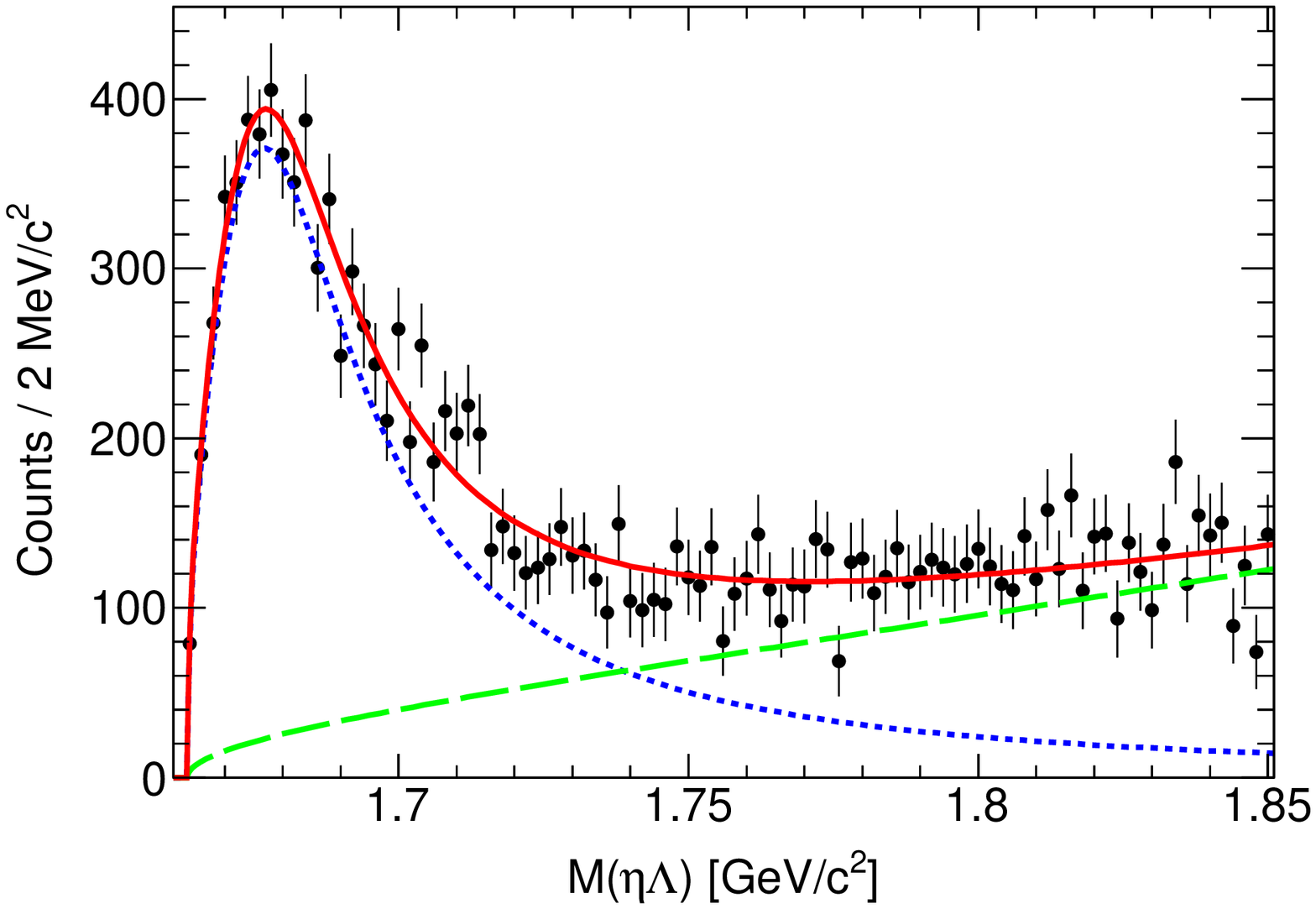}
  \end{minipage}\par\vspace{-0.5\baselineskip}
  \begin{minipage}[!htb]{1.0\linewidth}
      \includegraphics[height=0.75\textwidth,width=1.0\textwidth]{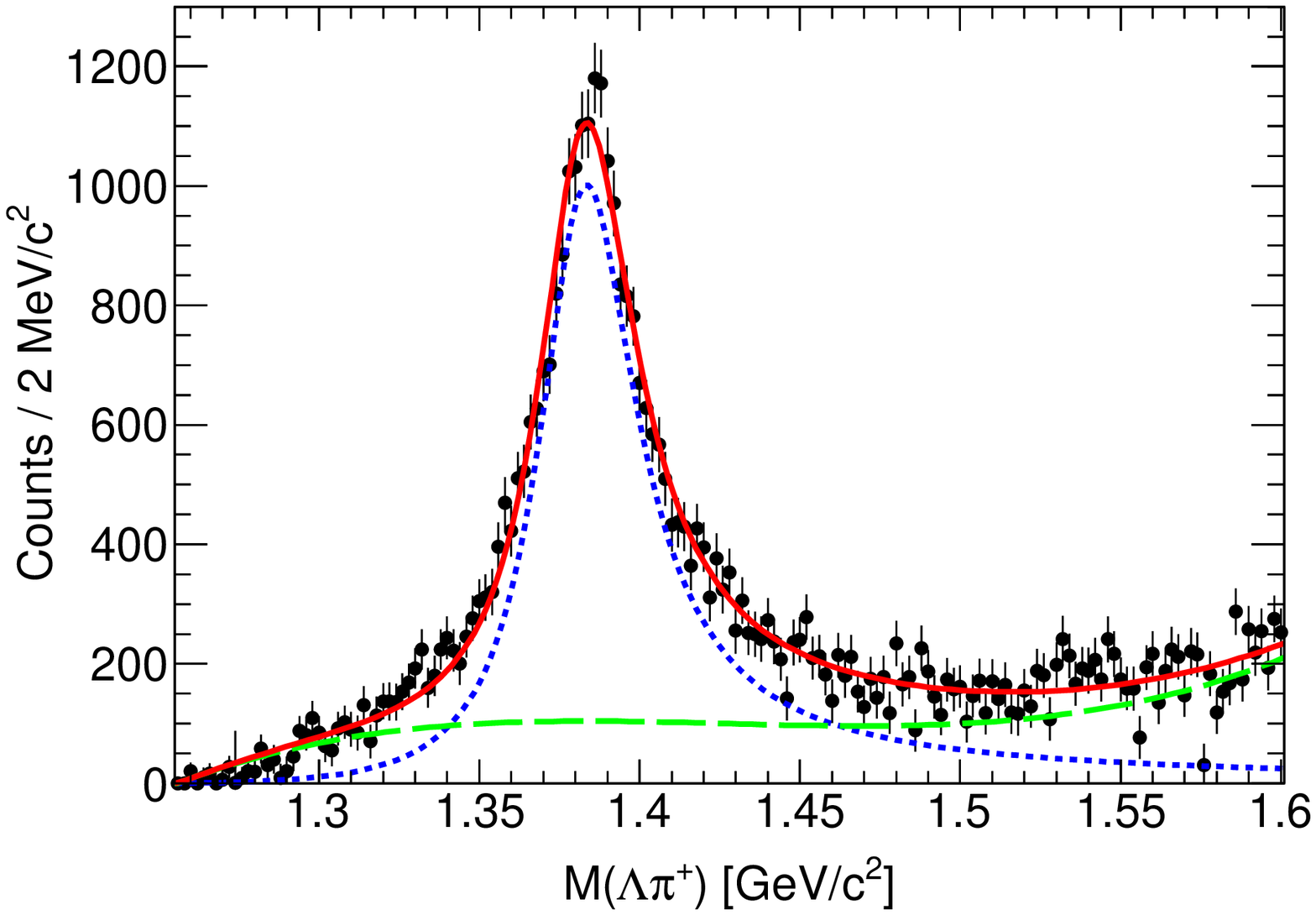}
  \end{minipage}\hspace{\fill}
  \caption
      {
        Fits to the $\lambdac$ yield in the $M(\eta\Lambda)$ (top) and $M(\Lambda\piplus)$ (bottom) spectra.
        The curves indicate the fit results: the total PDFs (solid red), the signal PDFs modeled with a relativistic Breit-Wigner function (dashed blue), and the background PDFs (long-dashed green).
      }
      \label{fig:sub_fit}
\end{figure}
In order to extract yields for the $\el$ and $\lpi$ contributions to inclusive $\elpi$ decays, we fit the $M(\eta\Lambda\pi^{+})$ mass distributions, and extract $\lambdac$ signal yields, for every 2 $\MeV$ bin of the $M(\eta\Lambda)$ and $M(\Lambda\piplus)$ distributions.
The same form of PDF described in Sec.~\ref{sec:branching} is used to fit the $M(\eta\Lambda\pi^{+})$ mass spectrum, and the PDF parameters for each mass bin are obtained in the same way for the fit of each Dalitz plot bin in Sec.~\ref{sec:branching}.
The $\lambdac$ yields as a function of $M(\eta\Lambda)$ and $M(\Lambda\piplus)$ are shown in Fig.~\ref{fig:sub_fit}.
The $\Lambda(1670)$ and $\Sigma(1385)^{+}$ resonances are clearly seen in Fig.~\ref{fig:sub_fit}(top) and (bottom), respectively.
This is the first observation of the $\Lambda(1670)$ in $\elpi$ decays. \par

To extract the signal yields for the two resonant decay modes, binned least-$\chi^{2}$ fits are performed to the $M(\eta\Lambda)$ and $M(\Lambda\piplus)$ spectra shown in Fig.~\ref{fig:sub_fit}.
 For the signal modeling, we use an $S$-wave relativistic partial width Breit-Wigner (BW) for the $\Lambda(1670)$ and a corresponding $P$-wave BW for the $\Sigma(1385)^{+}$:

\begin{equation}  \label{eq:breit-wigner1}
  \begin{aligned}
    \frac{dN}{dm} &\propto \frac{m\Gamma(m)}{\left(m^{2}-m_0^{2}\right)^{2}+m_{0}^{2}\left(\Gamma(m) + \Gamma_{\rm others }\right)^{2}},\\
  \end{aligned}
\end{equation}
with
\begin{equation} \label{eq:breit-wigner2}
  \begin{aligned}
    \Gamma(m)&=\Gamma_{0}\frac{m_{0}}{m}\left(\frac{q}{q_{0}}\right)^{2L+1}F(q),\\
  \end{aligned}
\end{equation}
where $m$, $m_{0}$ and $L$ are the invariant mass, the nominal mass and the decay angular momentum, respectively, and $q$ and $q_{0}$ are the center-of-mass momenta corresponding to $m$ and $m_{0}$, respectively.
Here $\Gamma(m)$ is the partial width for $\Lambda(1670) \ra \eta\Lambda$ or $\Sigma(1385)^{+} \ra \Lambda\piplus$ and $\Gamma_0 = \Gamma(m_{0})$ is a floating parameter in the fit. 
 The contribution $\Gamma_{\rm others}$, which indicates the sum of the partial widths for the other decay modes, is fixed to $25~\rm MeV$ for $\Lambda(1670)$ and $5~\rm MeV$ for $\Sigma(1385)^{+}$~\cite{pdg}. 
Unlike the $\Sigma(1385)^{+}$, the branching fractions for $\Lambda(1670)$ decays are not well determined~\cite{pdg}, we select 25 $\rm MeV$ as the nominal value for $\Gamma_{\rm others}$. 
A systematic uncertainty from the fixed value of $\Gamma_{\rm others}$ is calculated by changing this value over a wide range, 15 to 32 $\rm MeV$.
In~Eq.~\eqref{eq:breit-wigner2}, the Blatt-Weisskopf centrifugal barrier factor $F(q)$ is $1$ for $S$ wave and $(1+R^{2}q_{0}^{2})/(1+R^{2}q^{2})$ for $P$ wave, with $R={3.1~\rm GeV}^{-1}$~\cite{bw_formfactor}.
The detector resolution for $\Lambda(1670)$ is not included in the signal PDF because the detector response function is not a simple Gaussian near threshold.
The effect is small and is treated as a systematic uncertainty in the measurement.
On the other hand, for $\Sigma(1385)^{+}$ the relativistic Breit-Wigner function is convolved with a Gaussian with $\sigma = 1.39~\MeV$ to form the signal PDF.
This $\sigma$ value is determined from a MC simulation of detector responses.
To represent the background to the $\Lambda(1670)$ signal, we use a function with a threshold: $\sqrt{m-m_{\Lambda\eta}}\left[p_{0}+p_{1}\left(m-m_{\Lambda\eta}\right)\right]$, where $p_{0}$ and $p_{1}$ are free parameters and $m_{\Lambda\eta}$ is the sum of the masses of $\Lambda$ and $\eta$.
In the case of the $\Sigma(1385)^{+}$ fit, a third-order Chebyshev polynomial function is used to represent background.
The $\chi^{2}/\rm ndf$ of the $\Lambda(1670)$ and $\Sigma(1385)^{+}$ fits are 90.3/90 and 194/167, respectively.
We calculate the corresponding reconstruction efficiencies of $\el$ and $\lpi$ decays from a MC simulation. 
The extracted yields from the fits in Fig.~\ref{fig:sub_fit} are divided by the reconstruction efficiencies and the results are summarized in Table~\ref{tbl:yields}. The branching fractions relative to $\pkpi$ decay are summarized in Table~\ref{tbl:branching_fraction}. \par

 From the fit results, we also determine masses and widths ($\Gamma_{\rm tot} = \Gamma_{0} + \Gamma_{\rm others}$) of the $\Lambda(1670)$ and $\Sigma(1385)^{+}$ as summarized in Table~\ref{tbl:mass_width}.
 Changes in efficiency over the $M(\eta\Lambda)$ and $M(\Lambda\pi^{+})$ distributions are not considered because their effect is negligible as described in Sec.~\ref{sec:systematic}.
 The results obtained for the $\Sigma(1385)^+$ are consistent with previous measurements~\cite{pdg}.
 For the $\Lambda(1670)$, the mass and width have not been previously measured directly from a peaking structure in the mass distribution.
 The values that we obtain fall within the range of the partial wave analyses of the $\bar{K}N$ reaction~\cite{l1670_kamano, l1670_zhang}.

\begin{table}[!htb]
  \caption
      {\label{tbl:mass_width}
        Results for mass and width of the $\Lambda(1670)$ and $\Sigma(1385)^{+}$.
        The first and second uncertainties are statistical and systematic, respectively.
      }
  \begin{ruledtabular}
    \begin{tabular}{lcc}
      \textrm{Resonances}&
      \textrm{Mass} $[{\rm MeV}/c^{2}]$ &
	  \textrm{Width} $[{\rm MeV}]$ \\     
      \colrule
      $\Lambda(1670)$ & $1674.3 \pm 0.8 \pm 4.9$ & $36.1 \pm 2.4 \pm 4.8$ \\
      $\Sigma(1385)^{+}$ & $1384.8 \pm 0.3 \pm 1.4$ & $38.1 \pm 1.5 \pm 2.1$ \\
    \end{tabular}
  \end{ruledtabular}
\end{table}

\section{Systematic Uncertainty}
\label{sec:systematic}
The systematic uncertainties for the $\elpi$, $\eta\Sigma^{0}\piplus$, and $p\Kminus\piplus$ efficiency-corrected yields are listed in Table~\ref{tbl:sys_err}.
A study is performed based on a $D^{*+} \ra D^{0}\piplus(D^{0} \ra \Kminus\piplus)$ control sample for $\pi\/K$ identification and on the $\Lambda \ra p\pi^{-}$ decay for the proton identification to give corrections for the reconstruction efficiencies and to estimate the systematic uncertainties due to the PID selection.
Conservatively, all PID systematic uncertainties are considered to be independent when calculating the relative branching fractions to the $\pkpi$ channel.
The systematic uncertainty due to $\Lambda$ reconstruction is determined from a comparison of yield ratios of $B\rightarrow\Lambda\bar{\Lambda}K^{+}$ with and without the $\Lambda$ selection cut in data and MC samples.
The weighted average of the difference between data and MC samples over the momentum range is assigned as the systematic uncertainty.
A 3.0\% systematic uncertainty attributed to $\eta$ reconstruction is assigned by comparing the MC and data ratios of $\pi^{0}$ reconstruction efficiency for $\eta \rightarrow 3\pi^{0}$ and $\eta \rightarrow \pi^{+} \pi^{-} \pi^{0}$ decays~\cite{eta_uncertainty}.
The binning over the Dalitz plots is varied from $10 \times 5$ to $ 6\times 4$ and the differences in the results are taken as a systematic uncertainty.
Unlike the $\elpi$ and $\pkpi$ channels that are analyzed in a model-independent way, the efficiency of the $\sepi$ decay mode depends on its substructure.
To estimate the effect of possible substructures in the $\sepi$ decay, efficiencies of $\lambdac \ra \eta\Sigma(1385)^{+} \ra \eta\Sigma^{0}\piplus$ and $\lambdac \ra \Sigma^{0} a_{0}(980)^{+} \ra \eta\Sigma^{0}\piplus$ modes are compared to that of the nonresonant decay mode of $\sepi$ which is used to correct the yield and the larger difference is taken as systematic uncertainty.
The systematic uncertainty due to the background PDF modeling is estimated by changing the polynomial function from third order to fourth order.

In addition, the systematic uncertainties from the subdecay mode analysis that are not in common with the $\elpi$ decay channel are summarized in Table~\ref{tbl:sys_err_sub} and described below.
In order to estimate the systematic uncertainty due to $\Gamma_{\rm others}$, its value in the $\Lambda(1670)$ $\left(\Sigma(1385)^{+}\right)$ fit is varied from 15 to 32 (2 to 8) $\rm MeV$ and the maximum difference is taken as the systematic uncertainty.
The ranges of $\Gamma_{\rm others}$ conservatively cover the branching fractions of $\Lambda(1670)$ and $\Sigma(1385)^{+}$ decays in Ref.~\cite{pdg} and the $q$ dependence of $\Gamma_{\rm others}$ is negligible compared to this systematic uncertainty.
In the $M(\eta\Lambda)$ spectrum, the mass resolution varies from 0 to 2 $\MeV$ depending on mass; thus, two fits are performed by setting the mass resolution to 1 $\MeV$ and 2 $\MeV$, and the maximum difference is assigned as a systematic uncertainty.
For the $M(\Lambda\piplus)$ spectrum, we increase the detector resolution by 20\% and the resultant change is taken as a systematic uncertainty.
The systematic uncertainties from the background PDF modeling are estimated by fits with fixed shapes of background PDFs, which are determined by MC simulations including known background sources such as $\Lambda_{c}^{+} \rightarrow a_{0}(980)^{+} \Lambda$, nonresonant, and $\Lambda_{c}^{+} \rightarrow \eta \Sigma(1385)^{+}$ ($\Lambda_{c}^{+} \rightarrow \Lambda(1670)\pi^{+}$) decays in the $M(\eta\Lambda)$ ($M(\Lambda\pi^{+})$) spectrum.
In order to consider systematic uncertainties related to angular distributions of $\Lambda(1670)$ and $\Sigma(1385)^{+}$, the efficiencies in 10 bins of helicity angle are calculated and the largest efficiency differences between any efficiency in the helicity angle bin and the efficiency used to correct the yields are taken as systematic uncertainties.
It is possible that the results for the $\Lambda(1670)$ and $\Sigma(1385)^{+}$ can be affected by another resonant channel, $\Lambda_{c}^{+} \rightarrow a_{0}(980)^{+}\Lambda$. 
To estimate the interference effect with $a_{0}(980)^{+}$, we apply an additional $a_{0}(980)^{+}$ veto selection, removing events from $0.95$ to $1.02~{\rm GeV}/c^{2}$ of $M(\eta\pi^{+})$, to the $M(\eta\Lambda)$ and $M(\Lambda\pi^{+})$ distributions and subsequently repeat the fits.
By comparing the fit results with and without the $a_{0}(980)^{+}$ requirement, we determine the systematic uncertainties in the masses and widths.
For the efficiency-corrected yields, the expected yields calculated on the assumption that there is no interference effect are compared to the nominal values. 
Since the centrifugal barrier factor~\cite{bw_formfactor} is a model-dependent parameter, it has a sizeable uncertainty. 
Varying the parameter $R$ by $\pm 0.3$ GeV$^{-1}$, fits are performed to estimate the systematic uncertainty. 
We also estimate a systematic uncertainty from binning of $M(\eta\Lambda)$ and $M(\Lambda\pi^{+})$ distributions by changing the bin widths to $1~{\rm MeV}/c^{2}$.
 
The systematic uncertainties for the mass and width measurements are listed in Table~\ref{tbl:sys_err_mass_width}. 
In the same way as described above, the systematic uncertainties from the PDFs and the binning of the $\Lambda(1670)$ and $\Sigma(1385)^{+}$ fits are estimated. 
The absolute mass scaling is determined by comparing the measured mass of $\Lambda_{c}^{+}$ with that in Ref.~\cite{pdg}, and it is considered as a systematic uncertainty.
 To estimate the systematic uncertainty due to the $M(\eta\Lambda)$- and $M(\Lambda\pi^{+})$-dependent reconstruction efficiencies, we apply reconstruction efficiency corrections to the $M(\eta\Lambda)$ and $M(\Lambda\pi^{+})$ spectra.
For the corrections, we calculate the mass dependencies of these efficiencies by MC simulation. 
They are found to vary between 0.068 and 0.070 for $M(\eta\Lambda)$ and between 0.069 and 0.071 for $M(\Lambda\pi^{+})$, and in both cases the behavior is nearly flat. 
The mass spectra are divided by these efficiencies.
Differences in fit results with and without the efficiency corrections are negligible compared to these other systematic sources as listed in Table~\ref{tbl:sys_err_mass_width}.

\begin{table}[!htb]
  \caption
      {
        \label{tbl:sys_err}
        Summary of the systematic uncertainties (in \%) in the efficiency-corrected yields for the $\elpi$, $\sepi$ and $\pkpi$ channels.
      }
      \begin{ruledtabular}
        \begin{tabular}{lccc}
          {Source} & {$\eta\Lambda\pi^{+}$} & {$\eta\Sigma^{0}\piplus$ } & {$p\Kminus\piplus$} \\
          \colrule
          {PID} & 1.1 & 1.1 & 1.4 \\
          {$\Lambda$ reconstruction} & 2.8 & 2.8 & -  \\
          {$\eta$ reconstruction} & 3.0 & 3.0 & -  \\
          {Dalitz plot binning} & 1.3 & - & 0.7 \\
          {Intermediate states} & - & 1.3 & - \\
          {Background PDF} & 0.6 & 0.8 & 0.4 \\
          {MC statistics} & 0.2 & 0.2 & 0.1 \\
          {$\mathcal{B}_{\rm PDG}$} & 0.9 & 0.9 & - \\
          \colrule
          {Total} & 4.6  & 4.6 &  1.6  \\
        \end{tabular}
      \end{ruledtabular}
\end{table}

\begin{table}[!htb]
  \caption
      {
        \label{tbl:sys_err_sub}
        Summary of the systematic uncertainties (in \%) in the efficiency-corrected yields for the $\el$ and $\lpi$ channels that are not shared with $\elpi$ channel.
        The last row gives the total systematic uncertainty (and including the common sources, which are $\Lambda$ reconstruction and $\eta$ reconstruction, in Table.~\ref{tbl:sys_err}).
      }
      \begin{ruledtabular}
        \begin{tabular}{lcc}
          {~~~Source~~~} & {~~~$\Lambda(1670)$~~~} & { ~~~$\Sigma(1385)^{+}$~~~ }\\
          \colrule   
          {PID} & 1.0 & 1.1 \\       
          {$\Gamma_{{\rm others}}$} & 2.1 & 1.4  \\
          {Detector resolution} & 1.6 & 1.8  \\
          {Background modeling} & 11.6 & 2.8\\
          {Efficiency variation} & \multirow{2}{*}{1.8} & \multirow{2}{*}{5.5}   \\
          {over helicity angle} &  &    \\
          {Centrifugal barrier} & - & 0.7   \\
          {$\mathcal{B}_{\rm PDG}$} & 0.9 & 2.0 \\
          {MC statistics} & 0.2 & 0.2 \\
          {Bin width} & 1.7 & 1.2 \\
		  {Interference with $a_{0}(980)^{+}$} & 1.5 & 0.6 \\
          \colrule
          {Total} & 12.4 (13.0)  & 7.1 (8.2)\\
        \end{tabular}
      \end{ruledtabular}
\end{table}

\begin{table*}[!htb]
  \caption
      {
        \label{tbl:sys_err_mass_width}
        Summary of the systematic uncertainties in the masses and widths for the $\Lambda(1670)$ and $\Sigma(1385)^{+}$.  
      }
      \begin{ruledtabular}
        \begin{tabular}{lcccc}
         \multirow{2}{*}{Source} & \multicolumn{2}{c}{$\Lambda(1670)$} & \multicolumn{2}{c}{$\Sigma(1385)^{+}$} \\
                                 & {~~~Mass $[{\rm MeV}/c^{2}]$~~~} & { ~~~Width $[{\rm MeV}]$~~~ } & {~~~Mass $[{\rm MeV}/c^{2}]$~~~} & { ~~~Width $[{\rm MeV}]$~~~ } \\
          \colrule          
          {$\Gamma_{{\rm others}}$} & $3.6$ & $2.0$ & $0.3$ & $0.8$  \\
          {Detector resolution} & $0.4$ & $0.5$ & $0.0$ & $0.8$  \\
          {Background modeling} & $0.9$ & $3.9$ & $0.4$ & $1.5$  \\
          {Centrifugal barrier} & - & - & $0.1$ & $0.6$ \\
          {Bin width} & $0.0$ & $0.8$ & $0.1$ & $0.7$  \\
          {Mass scaling} & $0.2$ & - & $0.2$ & -  \\
          {Efficiency correction} & $0.1$ & $0.0$ & $0.1$ & $0.2$  \\
		  {Interference with $a_{0}(980)^{+}$} & $3.1$ & $1.5$ & $1.3$ & $0.2$ \\
          \colrule
          {Total} & $4.9$ & $4.8$ & $1.4$ & $2.1$ \\
        \end{tabular}
      \end{ruledtabular}
\end{table*}

\section{Summary}
We analyze the $\eta\Lambda\pi^{+}$ final state to study $\lambdac$ decays using the full data set of $980~\mathrm{fb^{-1}}$ at or near the $\Upsilon(nS)$ resonances collected by the Belle detector.
Two new decay modes of the $\lambdac$ baryon, $\sepi$ and $\el$, are observed for the first time, and their branching fractions are measured relative to that of the $\pkpi$ decay mode.
In addition, the branching fractions for $\elpi$ and $\lpi$, which were reported previously by CLEO~\cite{CLEO_br} and by BES\RN{3}~\cite{BES_br}, are measured with much improved precision.
The results are 
\begin{gather*}
\mfrac{\belpi}{\bpkpi} = \resultelpiratio, \\
\mfrac{\bsepi}{\bpkpi} = \resultsepiratio, \\
		\mfrac{\bel}{\bpkpi}{~~~~~~~~~~~~~~~~~~~~~~~~~~~~~~~~~~~} \\
{~~~~~~~~~~~~~~~~~~~~~~~~~~} = \resultelratio, \\
{\rm and~~~~~~~~~~~~~~~~~~~~~~~~~~~~~~~~~~~~~~~~~~~~~~~~~~~~~~~~~~~~~~~~~~~~~~~~~~~~~~~} \\
\mfrac{\blpi}{\bpkpi} = \resultlpiratio,
\end{gather*}
where the uncertainties, here and below, are statistical and systematic, respectively.
Assuming $\bpkpi$ = $ (6.28 \pm 0.32)\%$ \cite{pdg}, the absolute branching fractions are 
\begin{gather*}
\belpi = \resultelpiabs, \\
\bsepi = \resultsepiabs, \\
\bel {~~~~~~~~~~~~~~~~~~~~~}\\
{~~~~~~~~~~~~~~~~~}= \resultelabs,\\
{\rm and~~~~~~~~~~~~~~~~~~~~~~~~~~~~~~~~~~~~~~~~~~~~~~~~~~~~~~~~~~~~~~~~~~~~~~~~~~~~~~~} \\
\blpi = \resultlpiabs,
\end{gather*} 
where the third uncertainty is from $\mathcal{B}(\pkpi)$.
The measurements of $\belpi$ and $\blpi$ are the most precise results to date and agree with earlier results reported by CLEO~\cite{CLEO_br} and by BES\RN{3}~\cite{BES_br}.
In our study, the mass and width of the $\Lambda(1670)$ and $\Sigma(1385)^{+}$ are also determined to be 
\begin{gather*}
  m_{0}(\Lambda(1670)) = \resultlm, \\
  \Gamma_{\rm tot}(\Lambda(1670)) = \resultlw, \\
  m_{0}(\Sigma(1385)^{+}) = \resultsm, \\
{\rm and~~~~~~~~~~~~~~~~~~~~~~~~~~~~~~~~~~~~~~~~~~~~~~~~~~~~~~~~~~~~~~~~~~~~~~~~~~~~~~~} \\
  \Gamma_{\rm tot}(\Sigma(1385)^{+}) = \resultsw. \\
\end{gather*}
 These are the first measurements of the $\Lambda(1670)$ mass and width that are determined directly from a peaking structure in the mass distribution.



\begin{acknowledgments}
We thank the KEKB group for the excellent operation of the
accelerator; the KEK cryogenics group for the efficient
operation of the solenoid; and the KEK computer group, and the Pacific Northwest National
Laboratory (PNNL) Environmental Molecular Sciences Laboratory (EMSL)
computing group for strong computing support; and the National
Institute of Informatics, and Science Information NETwork 5 (SINET5) for
valuable network support.  We acknowledge support from
the Ministry of Education, Culture, Sports, Science, and
Technology (MEXT) of Japan, the Japan Society for the 
Promotion of Science (JSPS), and the Tau-Lepton Physics 
Research Center of Nagoya University; 
the Australian Research Council including grants
DP180102629, 
DP170102389, 
DP170102204, 
DP150103061, 
FT130100303; 
Austrian Science Fund (FWF);
the National Natural Science Foundation of China under Contracts
No.~11435013,  
No.~11475187,  
No.~11521505,  
No.~11575017,  
No.~11675166,  
No.~11705209;  
Key Research Program of Frontier Sciences, Chinese Academy of Sciences (CAS), Grant No.~QYZDJ-SSW-SLH011; 
the  CAS Center for Excellence in Particle Physics (CCEPP); 
the Shanghai Pujiang Program under Grant No.~18PJ1401000;  
the Ministry of Education, Youth and Sports of the Czech
Republic under Contract No.~LTT17020;
the Carl Zeiss Foundation, the Deutsche Forschungsgemeinschaft, the
Excellence Cluster Universe, and the VolkswagenStiftung;
the Department of Science and Technology of India; 
the Istituto Nazionale di Fisica Nucleare of Italy; 
National Research Foundation (NRF) of Korea Grant
Nos.~2016R1\-D1A1B\-01010135, 2016R1\-D1A1B\-02012900, 2018R1\-A2B\-3003643,
2018R1\-A6A1A\-06024970, 2018R1\-D1A1B\-07047294, 2019K1\-A3A7A\-09033840,
2019R1\-I1A3A\-01058933;
Radiation Science Research Institute, Foreign Large-size Research Facility Application Supporting project, the Global Science Experimental Data Hub Center of the Korea Institute of Science and Technology Information and KREONET/GLORIAD;
the Polish Ministry of Science and Higher Education and 
the National Science Center;
the Ministry of Science and Higher Education of the Russian Federation, Agreement 14.W03.31.0026; 
University of Tabuk research grants
S-1440-0321, S-0256-1438, and S-0280-1439 (Saudi Arabia);
the Slovenian Research Agency;
Ikerbasque, Basque Foundation for Science, Spain;
the Swiss National Science Foundation; 
the Ministry of Education and the Ministry of Science and Technology of Taiwan;
and the United States Department of Energy and the National Science Foundation.
J.Y. Lee and S.K. Kim were supported by NRF Grant No. 2016R1A2B3008343.
S.B. Yang acknowledges support from NRF Grant No. 2018R1A6A3A01012138.
Y. Kato is supported by MEXT KAKENHI Grant No. JP19H05148.
\end{acknowledgments}


\begin{thebibliography}{30}
\bibitem{charge-conjugate}
Unless otherwise stated, charge-conjugate modes are always implied throughout this paper.
\end{thebibliography}

\end{document}